\documentclass[useAMS,usenatbib]{mn2e}

\usepackage{lscape,graphicx,pifont}
\usepackage{times}
\usepackage{longtable}
\usepackage{natbib}
\usepackage{color}

 \DeclareGraphicsExtensions{.png,.pdf,.eps,.jpg,.tiff}

\newcommand{\Msun}{\mbox{\,M$_\odot$}}

\newcommand{\Rsun}{\mbox{\,R$_\odot$}}
\newcommand{\vunit}{\mbox{\,km\,s$^{-1}$}}
\newcommand{\mic}{\mbox{$\,\mu$m}}
\newcommand{\pion}[2]{{#1}\,{\sc {#2}}}

\newcommand{\Teff}{\mbox{$T_{\rm eff}$}}

\newcommand{\ltsimeq}{\raisebox{-0.6ex}{$\,\stackrel
        {\raisebox{-.2ex}{$\textstyle <$}}{\sim}\,$}}

\newcommand{\spitzer}{\mbox{\it Spitzer Space Telescope}}

\newcommand{\logg}{\mbox{$\log{g}$}}
\newcommand{\nucl}[2]{\mbox{$^{#1}${#2}}}

\newcommand{\tcrb}{\mbox{T~CrB}}

\title[Isotopic ratios in T~CrB]{Isotopic ratios in the red giant component 
of the recurrent nova T~Coronae Borealis}
\author[Ya. V. Pavlenko et al.]{Ya.V. Pavlenko$^{1,2}$\thanks{E-mail: yp@mao.kiev.ua},
A. Evans$^{3}$,
D. P. K. Banerjee$^{4}$,
T. R. Geballe$^{5}$,
U. Munari$^{6}$, \newauthor
R. D. Gehrz$^{7}$, 
C. E. Woodward$^{7}$, 
S. Starrfield$^{8}$ \\
\\
{$^1$}Main Astronomical Observatory, Academy of Sciences of the Ukraine, 
Golosiiv Woods, Kyiv-127, 03680 Ukraine\\ 
{$^2$}Centre for Astrophysics Research, University of Hertfordshire, College Lane, Hatfield, AL10 9AB, UK \\
{$^3$}Astrophysics Group, Lennard Jones Laboratory, Keele University, Keele, Staffordshire,  ST5 5BG, UK\\ 
$^4${Physical Research Laboratory, Navrangpura,  Ahmedabad, Gujarat 
380009, India}\\ 
$^{5}$Gemini Observatory, 670 N. Aohoku Place, Hilo, HI, 96720,
USA\\ 
$^6$INAF Astronomical Observatory of Padova, I-36012 Asiago (VI), Italy\\ 
$^7$Minnesota Institute for Astrophysics, School of Physics \& Astronomy,
116 Church Street SE, University of Minnesota, Minneapolis, MN 55455, USA\\ 
$^{8}$School of Earth and Space Exploration, Arizona State University, 
Box 871404, Tempe, AZ 85287-1404, USA\\ 
 }

\begin{document}

\date{Version of \today}

\pagerange{\pageref{firstpage}--\pageref{lastpage}} \pubyear{2020}

\maketitle

\label{firstpage}

\begin{abstract}
We report the determination of abundances and isotopic ratios for 
C, O and Si in the photosphere of the red giant component of the 
recurrent nova T~Coronae Borealis from new 2.284--2.402\mic\ and 
3.985--4.155\mic\ spectroscopy. Abundances and isotopic ratios in the 
photosphere may be affected by (i)~processes in the red giant interior
which are brought to the surface during dredge-up, (ii)~contamination of 
the red giant, either during the common envelope phase of the binary 
evolution or by material synthesised in recurrent nova eruptions, or a
combination of the two.
We find that the abundances of C, O and Si are reasonably consistent 
with the expected composition of a red giant after first dredge-up, 
as is the \nucl{16}{O}/\nucl{17}{O} ratio. 
The \nucl{28}{Si}/\nucl{29}{Si} ratio is found to 
be $8.6\pm3.0$, and that for \nucl{28}{Si}/\nucl{30}{Si} is $21.5\pm3.0$. The \nucl{12}{C}/\nucl{13}{C} 
ratio ($10\pm2$) is somewhat lower than expected for first dredge-up.
The \nucl{16}{O}/\nucl{18}{O} ratio ($41\pm3$)
is highly inconsistent with that expected either from red giant evolution 
($\sim550$) or from contamination 
of the red giant by the products of a nova thermonuclear runaway. In 
particular the C and O isotopic ratios taken in combination are a puzzle.
We urge confirmation of our results using spectroscopy at high resolution. 
We also encourage a thorough theoretical study of the effects on the 
secondary star in a recurrent nova system of contamination by ejecta
having anomalous abundances and isotopic ratios.
\end{abstract}

\begin{keywords}
nuclear reactions, nucleosynthesis, abundances ---
stars: AGB and post-AGB ---
stars: abundances ---
  stars: individual (\tcrb) --- 
 novae: cataclysmic variables ---
 infrared: stars 
 \end{keywords}


\section{Introduction}
\label{intro}

Recurrent novae (RNe) are cataclysmic variable (CV) systems that 
have been observed to 
undergo more than one nova outburst \citep[see e.g.][and references 
therein]{evans-asp}. They fall into two groups, having short 
($\sim1$~day) and long ($\sim1$~yr) orbital periods \citep{anupama},
the latter having red giant (RG) secondaries.
Long orbital periods demand RG secondaries so that they are
sufficiently large either to fill their Roche lobes, or to have large 
mass-loss rates in the form of winds.

Most RNe are believed to have white dwarf (WD) primaries with masses 
close to the Chandrasekhar limit. Like classical novae, RN eruptions 
are the result of a thermonuclear runaways (TNRs) on the surfaces of the WD
components. The recurrence time-scales for both 
types of RNe range from $\sim10$~yrs to $\sim80$~yrs 
\citep[see][and references therein]{schaefer}, which  
(in part at least) is an observational selection effect.

The RN \tcrb\ has undergone eruptions in 1866 and 1946,
and there are several indicators to 
suggest that another eruption is imminent \citep{schaefer,munari}.
The system has a long orbital period (227.67~d). The spectral
type of the RG is M3~III \citep{anupama}. 
Three studies have found that the mass of the WD is $\sim1.35$\Msun\
\citep[][]{shahbaz2,kennea,shara}, which is close to the Chandrasekhar 
limit. \citeauthor{shara} give  $2.1\times10^{-8}$\Msun~yr$^{-1}$
for the mass accretion rate.

\cite{evans4} recently presented NASA \spitzer\ \citep{spitzer,spitzer2}, 
{\it Stratospheric Observatory for Infrared Astronomy} 
\citep[SOFIA;][]{sofia} and other infrared (IR) observations of 
\tcrb. They reported the discovery of the SiO 8\mic\ 
fundamental transition in absorption, arising from both
the RG photosphere and its wind. 

Isotopic abundances of the elements in stars, supernovae and
meteors are good indicators of 
nucleo\-synthetic processes in the Galaxy. In particular, the isotopes 
of carbon, oxygen and silicon form under very specific conditions.
Knowledge of the isotopic compositions in stars is an important factor
in understanding Galactic chemical evolution \citep[see e.g.][]{romano}.
Using the first overtone of CO in absorption, \cite{evans4} determined that the
photospheric \nucl{12}{C}/\nucl{13}{C} ratio is $\ltsimeq9$ in the 
RG in \tcrb; this is considerably lower than the Solar System value of 89
\citep*[e.g.][]{asplund}.

The abundances of C, O and Si isotopes in the RG in \tcrb\ 
may differ from those in ``field'' RGs, as it might be
contaminated by material from the WD progenitor
(e.g. during the common envelope phase), or by ejecta from
RN eruptions. In the latter case for example, enhancements of \nucl{17}{O}
and \nucl{13}{C} might occur, depending on the nature of the 
RN eruption, but no significant enhancements of the Si isotopes should occur
unless the WD is of the ONe type \citep[see e.g.][]{starrfield09}.

Here we present near IR spectra at higher resolution 
than in \cite{evans4}, to isolate various isotopic 
bands. We  determine abundances and isotopic ratios for carbon, 
oxygen and silicon in the RG photosphere.
Following \cite{evans4}, we assume a reddening of $E(B-V) = 0.06$.

\section{Observations}

\label{gemini}

\begin{table*}
\caption{Observing log.\label{obslog}.}
 \begin{tabular}{cccccc}
UT Date & Airmass & Spectral range & Resolution & Integration & Telluric Standard  \\
YYYY-MM-DD &      &   ($\mu$m)     & ($R=\lambda/\Delta\lambda$) & time (s) & (airmass) \\ \hline
2019-04-10 & 1.02 & 2.284--2.346 & 19,000 & 320 & HIP 75178 (1.04) \\
2019-04-16 & 1.01 & 2.342--2.402 & 19,000 & 240 & HIP 75178 (1.01)\\
2019-04-19 & 1.20 & 4.065--4.155 & 22,000 & 640 & HIP 75178 (1.18)\\
2019-04-25 & 1.12 & 3.985--4.075 & 22,000 & 640 & HIP 83207 (1.07)\\ \hline
\end{tabular}
\end{table*}

Spectra of \tcrb\ were obtained in April 2019 at the Frederick C. Gillett 
Gemini North Telescope for program GN-2019A-FT-207. The facility near-infrared 
spectrograph GNIRS was  used with its $111~\ell$~mm$^{-1}$ grating, long focal
length camera, a $0.10''$ diameter slit, and in the normal 
stare/nod-along-slit mode.  
An observing log is provided in Table~\ref{obslog}. Two slightly overlapping 
spectral segments, each in the regions of the CO first overtone bands at 
2.3--2.4\mic\ and the SiO overtone bands at 4.00--4.15\mic, were observed. 
A telluric standard star near \tcrb\ (see Table~\ref{obslog}) was observed
either immediately before or immediately after \tcrb\ to ensure close 
airmass matches. 

Data reduction, utilising {\sc iraf} \citep{iraf1,iraf2} and 
{\sc figaro} \citep{figaro}, was  mostly standard, consisting of
flat-fielding, spectrum extraction, spike removal, combining spectra from the 
nodded positions, wavelength calibration (using telluric lines; accurate to 
0.00003\mic\ for the CO spectra and 0.00006\mic\ for  the SiO spectra), 
ratioing \tcrb\ spectra by the standard spectra, and stitching together the 
adjacent segments. In the shorter wavelength SiO spectrum the standard star's 
spectrum contains prominent \pion{H}{i} 5--4 and 14--6 absorption lines. 
To remove them from its spectrum prior to ratioing the same technique as 
described in \cite*{stencel} was employed.

There is evidence that the RG in \tcrb\ is irradiated by the WD
\citep[see][]{evans4} and we note that the orbital phase
on the date of the Gemini observation was 0.13,
with 0.0 corresponding to the RG near inferior conjunction 
(i.e., in front of the WD). We therefore ignore the effects
of irradiation of the RG by the WD.

\section{Synthetic spectra}
  
Theoretical synthetic spectra were computed using the program
WITA6 \citep*[see][for details]{pavl02,pavl03,pavl20a,ivanyuk}. We assumed 
local thermodynamic equilibrium (LTE), hydro-static equilibrium and a
one-dimensional (1D) model atmosphere without sources and sinks of
energy. Synthetic spectra were computed for RG
model atmospheres having effective temperatures 
$T_{\rm eff} = 3000 - 4000$~K, gravities \logg\ from 0 to 3 with an
increment $\Delta\logg =1.0$, and microturbulent velocity 3\vunit.  
The model atmosphere parameters for the \tcrb\ RG were determined in 
\cite{evans4} and \cite{woodward20}.

Molecular line lists were taken 
from the High-temperature Molecular Spectroscopic Database
\citep[HITEMP;][]{rothman}. Data for SiO and its isotopologues were 
taken from the Molecular line lists for Exoplanet and other Hot 
Atmospheres \citep[ExoMol;][]{barton} database.
The Einstein $A$ coefficients in the ExoMol line list for SiO 
are accurate to about 10--20\% (S. Yurchenko, private communication).
The uncertainties in $A$ will of course impact our abundance uncertainties,
but for consistency with the CO analsis and with previous work, 
we do not propagate these uncertainties into our final results. 
The dissociation energy of SiO was taken to be 8.26~eV
\citep[see e.g][]{cox}.
H$_2$O was accounted for using 
the ``BT2'' line list \citep{barber}. Computations, which included
all the most abundant isotopologues of CO and SiO, were performed for the 
1D SAM12 model atmospheres \citep{pavl03}. 

The best fits to the observed spectra were determined by minimising
the function $S= \sum_i^n s_i^2$, where 
\[ s_i = |(F^{\rm obs}_i - F^{\rm comp}_i)|, \: 
i=1, n \:\:; \] 
$F^{\rm obs}_i$, $F^{\rm comp}_i$, $n$ are the observed 
and computed fluxes at $n$ wavelengths, respectively. 
The minimisation parameter $S$ is found iteratively on the 3D 
grid of radial velocity sets, normalisation factors, and
broadening parameters. We also compute the errors in the
fits $\delta = \sum s_i/n$. 

\begin{figure}
   \centering
\includegraphics[width=8cm]{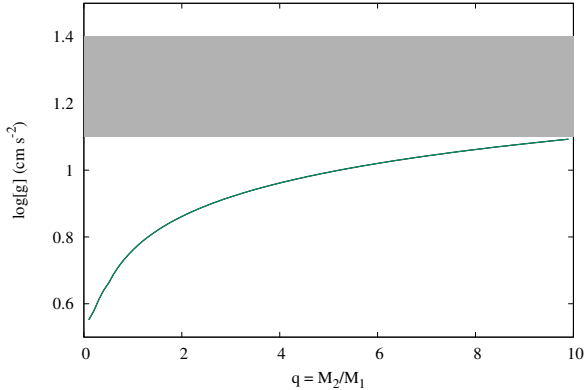}
\caption{Dependence of $\log{g}$ on the mass ratio $M_2/M_1$.
Solid curve corresponds to Equation~(\ref{g2}), and 
applies if the RG fills its Roche lobe. 
Shaded rectangle is typical $\log{g}$ for field RGs of spectral
type M3III. See text for details.\label{logg}}
  \end{figure}

  \begin{figure*}
   \centering
   \includegraphics[width=16cm,keepaspectratio]{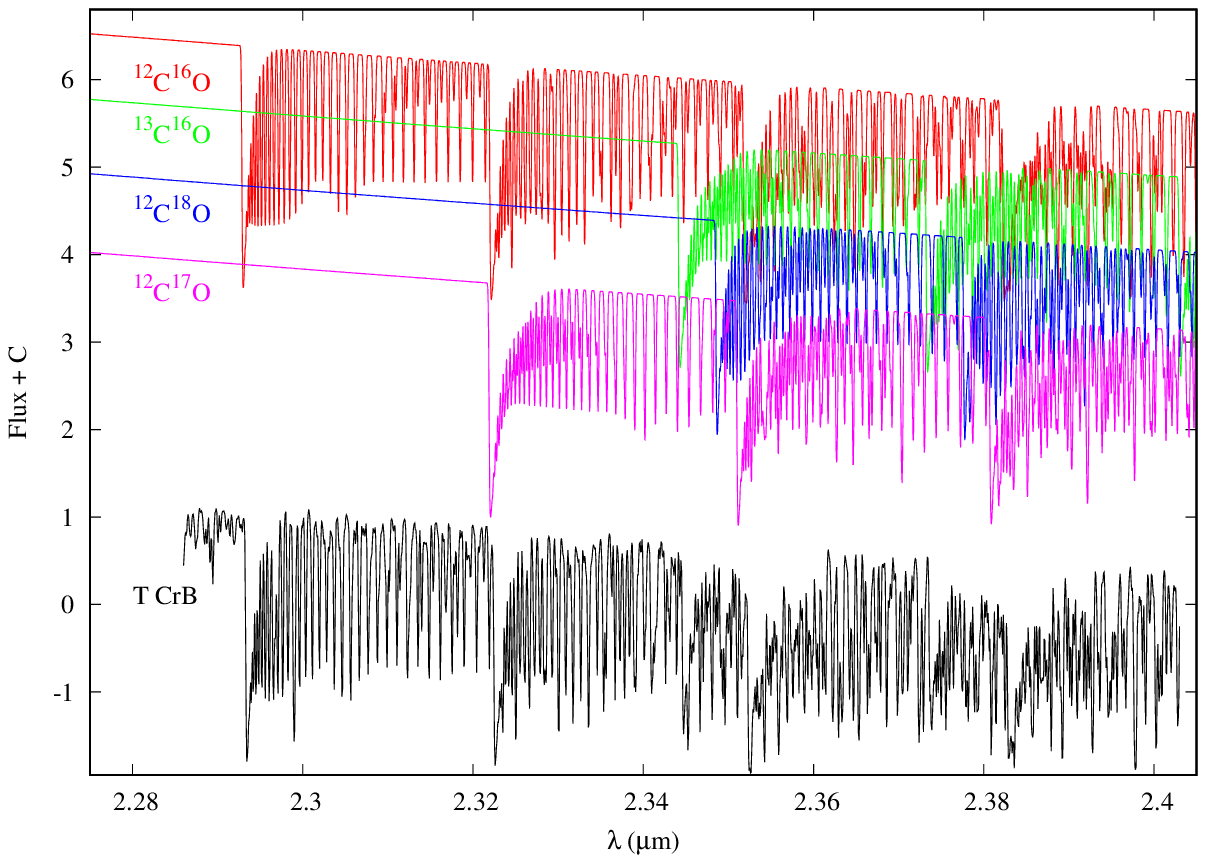}
\caption{Identification of the isotopic bands of CO.
\nucl{12}{C}\nucl{16}{O}: red;
\nucl{13}{C}\nucl{16}{O}: green;
\nucl{12}{C}\nucl{18}{O}: blue;
\nucl{12}{C}\nucl{17}{O}: magenta.
The \tcrb\ spectrum is also shown (black). 
Note that the flux scale is arbitrary. \label{ico}}
  \end{figure*}

To accelerate the iteration process, the theoretical spectrum 
is normalised to have flux 1.0 at the 
maximum of the spectral energy distribution (SED);
accordingly the fitted and observed spectra are displayed here
with flux in the range 0--1.
For each spectrum we determine the ``effective'' resolution $R$,
which differs from the nominal observational spectral resolution
because of the effects of macro/macroturbulent motions in the stellar 
atmosphere,  etc.
Synthetic spectra were computed for a model atmosphere 
with [\Teff,\logg, [Fe/H]] = [3600,1.0,0.0], convolved with 
$R=8$k, 9k, 10k, 11k, 12k, and fitted to the observed 
spectrum\footnote{Throughout we specify a model with the 
convention [\Teff,\logg,[Fe/H]].}. We find that the deduced
abundances have only a marginal dependence on $T_{\rm eff}$.
The optimal values of the atomic and
isotopic abundances for C and Si were finally determined using
the procedure described by \cite{pavl20b}.

\section{Estimating \logg\ for the RG}
\label{glog}
  
The strengths of the first overtone CO bands in stellar 
spectra depend on the effective temperature \Teff, the elemental and
molecular abundances, and gravity \logg. We have determined 
the effective temperature of the RG in \tcrb\ 
\citep[$\Teff = 3600$~K;][]{evans4} by fitting the 0.75--2.5\mic\ spectrum
(including the TiO bands), on the assumption that the abundances of 
metals and other molecules are solar. 

However, the carbon abundance 
in the RG in \tcrb\ may differ from solar. Furthermore, \logg\ is 
a poorly determined parameter, even for normal RGs. 
To estimate  \logg\ for the RG in \tcrb, 
we use the reasonably-well  constrained values for \tcrb,
namely the orbital period (227.67~d) and the mass of the WD (1.35\Msun;
see Section~\ref{intro}). 
There is strong evidence, in the form of ellipsoidal variations 
in the IR \citep{yudin93}, that the RG in \tcrb\ fills its Roche lobe.
For our present purposes we assume that it does so, and that the
volume of the RG equals the volume of its Roche lobe. The 
formulae summarised by \cite{king} give
\begin{eqnarray}
g & = & \frac{(GM_1)^{1/3}  (4\upi^2)^{2/3}}{P^{4/3}}
\:\: \frac{q}{(1+q)^{2/3} [f(q)]^2} \nonumber \\
    &   \simeq & 1.231 \left ( \frac{M_1}{1.35\Msun} \right )^{1/3} 
    \:\: \frac{q}{(1+q)^{2/3} [f(q)]^2}      \label{g2} 
\end{eqnarray}  
for the surface gravity $g$ (in cm~s$^{-2}$), where $q=M_2/M_1$ 
is the mass ratio, $M_1$ ($M_2$) being the mass of the WD
primary (RG secondary) star and $f(q) = R_2/a$ is an empirical 
function of $q$ that is determined by the ratio of the secondary 
Roche lobe radius ($R_2$) and the separation of the stars, $a$ 
\citep[see][and especially Section~2.3 therein,
for details]{king}. 
The dependence of $\log{g}$ on $q$ from Equation~(\ref{g2}) is shown in 
Fig.~\ref{logg}, in which it is clear that 
$0.6\ltsimeq\log{g}\ltsimeq1.1$ for a wide range of RG/WD mass ratios. 

Field M giants generally have masses $\sim1-2$\Msun\ and radii in the
range 40--100\Rsun\ for ``normal'' early- to mid-M-giants. The
corresponding \logg\ values (in cgs units) are of order 1.0 dex
for early-M and middle-M giants \citep[see][and references therein]{gray09};
the typical range of \logg\ for M giants is shown by the shaded region 
in Fig.~\ref{logg}. For the RG in the \tcrb\ system, $\log{g}$ is highly 
unlikely to be greater than this; indeed it is almost certainly smaller 
as the RG will likely be bloated as a result of irradiation by the WD.

We use $\log{g}=1$ and $\Teff = 3600$~K here.

\begin{table}
\caption{Wavelengths (in \mic, in vacuo) of the CO bandheads.
First overtone transitions are indicated by $\delta\upsilon$.
From https://www.gemini.edu/observing/resources/near-ir-resources/
spectroscopy/co-lines-and-band-heads. \label{COiso}.}
 \begin{tabular}{ccccc}
$\delta\upsilon$   & \nucl{12}{C}\nucl{16}{O} & \nucl{13}{C}\nucl{16}{O} &  
\nucl{12}{C}\nucl{18}{O}   & \nucl{12}{C}\nucl{17}{O} \\ \hline
2--0  &    2.2935 &   2.3448  &  2.3492  &   2.3226   \\
3--1  &    2.3227 &   2.3739  &  2.3783  &   2.3517  \\
4--2  &    2.3525 &   2.4037  &  2.4081  &   2.3815 \\
5--3  &    2.3829 &   2.4341  &  2.4385  &   2.4119  \\
6--4  &    2.4142 &   2.4652  &  2.4696   &   ---  \\
7--5  &    2.4461 &   2.4971  &   ---     &   --- \\
8--6  &    2.4787 &    ---    &   ---     &   --- \\
9--7  &    2.5122 &    ---    &   ---     &   --- \\ \hline
 \end{tabular}
\end{table}

\section{Abundances and isotopic ratios}

\subsection{Carbon abundance and the $^{12}$C/$^{13}$C ratio}
\label{Cisotp}

The carbon abundance and the C isotopic ratios were determined 
iteratively by fitting the first overtone CO bands. The band heads for 
the various CO isotopologues, together with the corresponding individual 
first overtone transitions, are shown in Fig.~\ref{ico} and listed in
Table~\ref{COiso}.

\begin{figure}
   \centering
      \includegraphics[width=9cm]{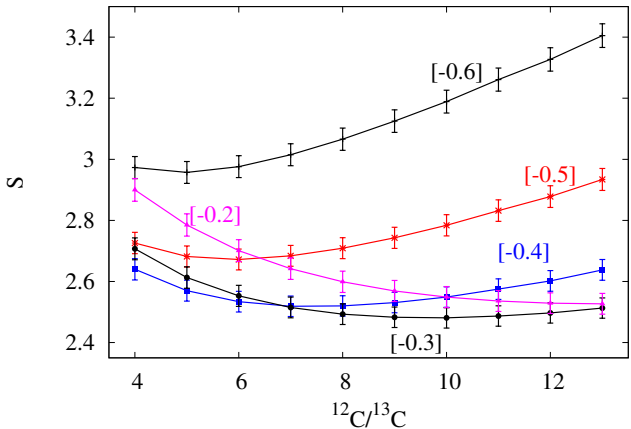}
\caption{Dependence of minimisation parameter $S$ on the adopted carbon 
abundance. Numbers in brackets are abundances of carbon measured 
relative to the solar carbon abundance $\log{N}({\mbox{C}})_{\odot} = 8.43$
on a scale having $\log{N}({\mbox{H}}) = 12$ \citep{asplund}.
\label{ccs}}
  \end{figure}

As a rule, lowering \logg\ increases the strength of 
CO bands in the synthetic spectra, so that the carbon abundance has
to be reduced to fit the observed spectrum. Furthermore,
the determined isotopic ratio \nucl{12}{C}/\nucl{13}{C}
is sensitive to changes in the carbon abundance: 
as $\log{N}(\mbox{C})$ is increased, the isotopic ratio 
\nucl{12}{C}/\nucl{13}{C} decreases ($N(\mbox{X})$ being defined
as the number of species X). 
This dependence  of \nucl{12}{C}/\nucl{13}{C}
on the C abundance arises because of the different strengths of 
the $^{12}$CO and $^{13}$CO features. The
$^{12}$CO bands are much the stronger, therefore the 
dependence on changes in the C abundance is much less pronounced. 
Hence, if we reduce the C abundance, there is a strong reduction
in the theoretical $^{13}$CO abundance, so we need to increase $^{13}$CO
by a {\em larger} amount than the factor by which $N(\mbox{C})$ was 
reduced to get the best fit to the observed spectrum.
This is illustrated in Fig.~\ref{ccs}, which shows the dependence of 
the fit parameter $S$ on the assumed carbon abundance.
We therefore recomputed the synthetic 
spectrum for each newly determined carbon abundance.

\begin{figure}
   \centering
      \includegraphics[width=9cm]{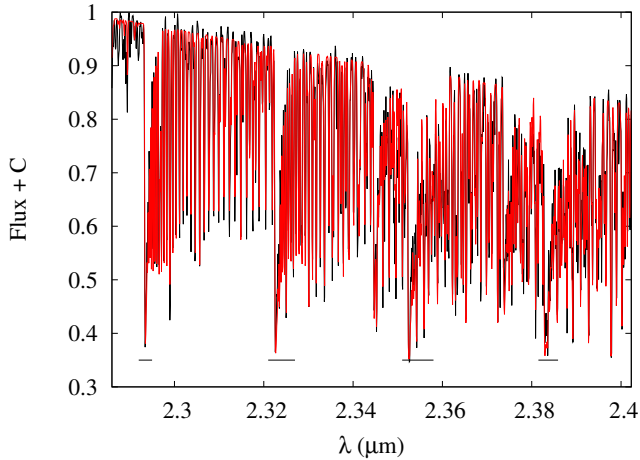}
\caption{The fit to the first overtone CO bands; 
the wavelength ranges highlighted by the short lines were omitted from
the analysis for reasons discussed in the text. \label{CO_SAT}}
  \end{figure}

We also note that the fluxes around the bandheads 
suffer from some saturation. To overcome this we determine the 
isotopic ratios and the C abundance using the wavelength range 
2.2860--2.4023\mic, omitting the data around the bandheads 
(see Fig.~\ref{CO_SAT}).
We found $\nucl{12}{C}/\nucl{13}{C}=10\pm2$,
$\log{N}(\mbox{C}) = 8.09\pm0.01$ (on a scale with $\log{N}(\mbox{H})=12$), 
i.e. $\mbox{[C]}=-0.3$, after three 
iterations\footnote{Including the
bandheads results in a value for \nucl{12}{C}/\nucl{13}{C} of $8.5\pm0.5$}.
This value of \nucl{12}{C}/\nucl{13}{C} is consistent with that obtained
by \citeauthor{evans4} (2019; $\nucl{12}{C}/\nucl{13}{C}\ltsimeq9$) using lower 
resolution spectra. 

\begin{figure}
   \centering
\includegraphics[width=8cm,keepaspectratio]{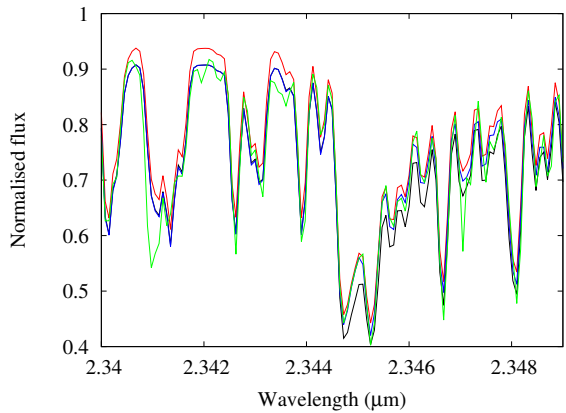}
\caption{Fit to the CO bands in the 2.340--2.349\mic\ region.
Black curve: \tcrb. 
Green curve: \nucl{12}{C}/\nucl{13}{C} = 7,
blue curve: \nucl{12}{C}/\nucl{13}{C} = 13.
Red curve: \nucl{12}{C}/\nucl{13}{C} = 10, as discussed in text 
(see Section~\ref{Cisotp}); 
this curve has been displaced upwards by $+0.03$ to 
emphasise the difference between it and the green curve, as these
are almost indistinguishable shortward of $\sim2.344$\mic.
\label{4710}}
  \end{figure}

We show in Fig.~\ref{4710} the fits to 
the data with \nucl{12}{C}/\nucl{13}{C} = 10.
Also shown are identical models (i.e. same $T_{\rm eff}$ etc.), 
but with \nucl{12}{C}/\nucl{13}{C} values that lie outside the 
deduced uncertainty of $\pm2$. This wavelength range covers the 
$\upsilon=0-2$ transition in \nucl{13}{C}\nucl{16}{O}, and 
demonstrates that the isotopic ratio deduced
provides a good fit to the data, whereas
values outside the uncertainty range do not.

\subsection{Oxygen isotopic ratios}

\subsubsection{\nucl{16}{\mbox{\rm O}}/\nucl{18}{\mbox{\rm O}}}

For the solar photosphere, $\nucl{12}{C}/\rm{C} = 0.989$, 
$\nucl{13}{C}/\rm{C} = 0.011$; the corresponding data for oxygen are
$\nucl{16}{O}/\rm{O} = 0.99762$, $\nucl{17}{O}/\rm{O} = 0.000379$
$\nucl{18}{O}/\rm{O} = 0.0020$ \citep{asplund}. 

Our spectra are of sufficient quality that we can determine 
\nucl{16}{O}/\nucl{18}{O} for the RG in \tcrb.
We first model the spectral range 2.28--2.42\mic, which includes all
the likely isotopologues of CO. To determine \nucl{18}{O}/\nucl{16}{O}, 
we fixed $\nucl{12}{C}/\nucl{13}{C}=10$ and
$\log{N}(\mbox{C}) = 7.92$, 
as found for the [3600/1.0/0.0] model atmosphere (see Section~\ref{Cisotp}). 

\begin{figure*}
   \centering
   \includegraphics[width=15cm,keepaspectratio]{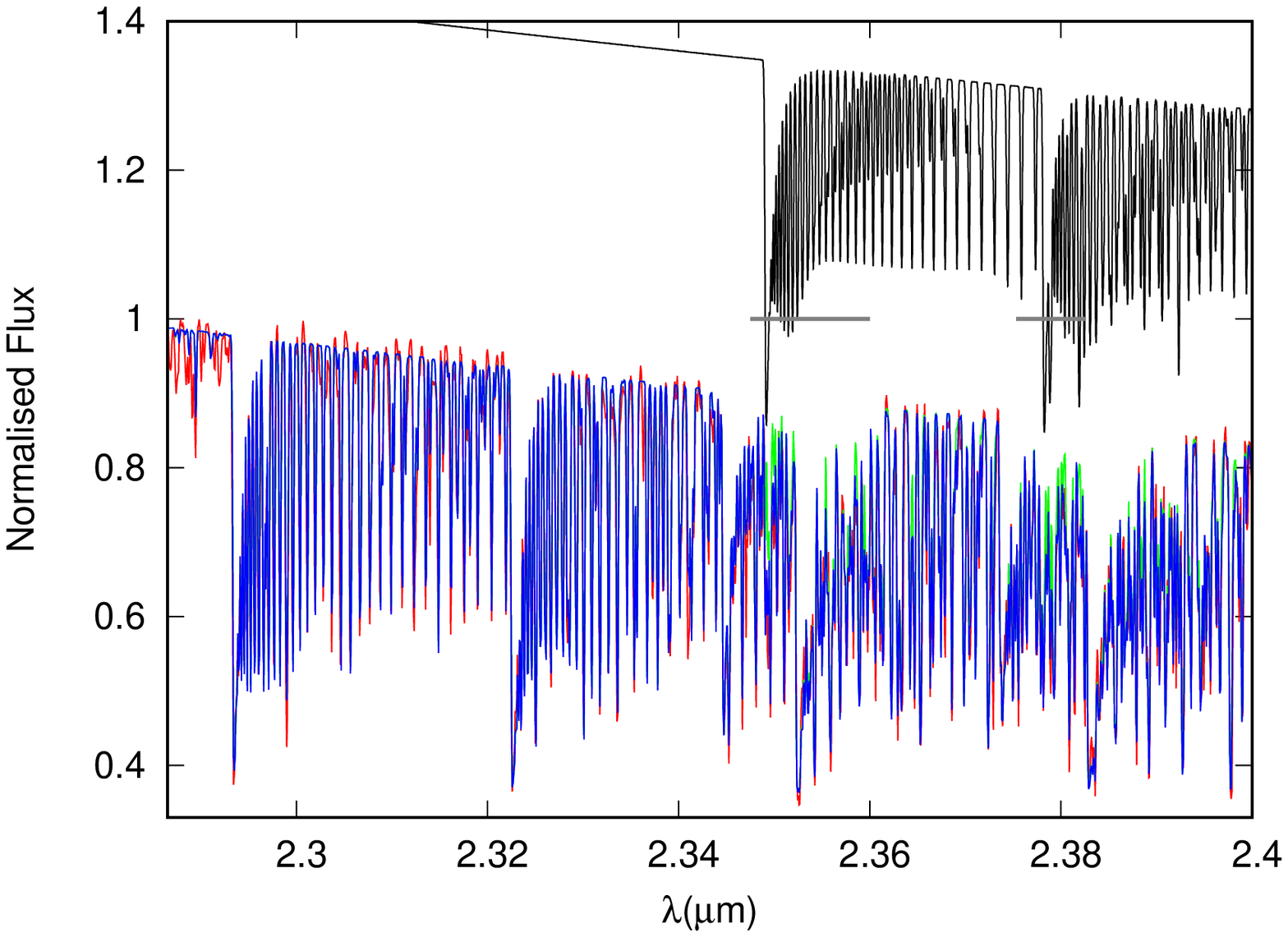}
   \includegraphics[width=15cm,keepaspectratio]{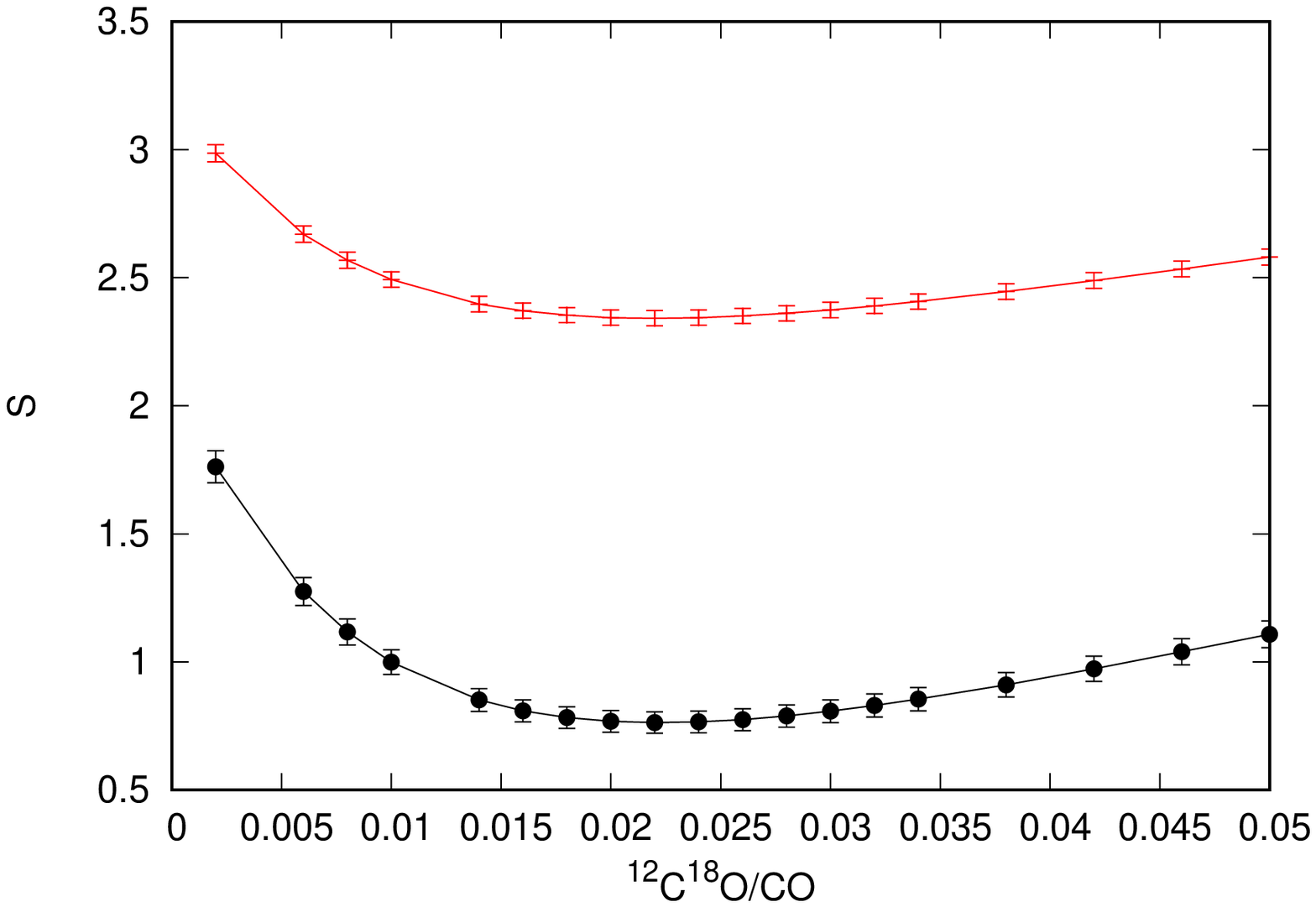}
\caption{Top: best fit to the observed spectrum (red), with (blue) and 
without (green) the \nucl{12}{C}\nucl{18}{O} bands. The 
\nucl{12}{C}\nucl{18}{O} bands are shown in black at the top.
Bottom: minimisation parameter $S$ for the fit of the synthetic 
spectra across bands of the first overtone CO 
over the full spectral region (red line) and across two
regions containing the \nucl{12}{C}\nucl{18}{O} bands (black line), 
highlighted in the top panel by the horizontal grey lines. 
See text for details. \label{18ccs}}
  \end{figure*}

The minimisation procedure is run for a set of synthetic spectra 
with $\nucl{12}{C}\nucl{18}{O}/\rm{CO}$ running from 0 to 0.12, with step 
size 0.002; in the first iteration we consider the entire spectral range, 
2.25--2.4\mic. The dependence of the minimisation parameter $S$ on the 
\nucl{12}{C}\nucl{18}{O}/\rm{CO} is shown in Fig.~\ref{18ccs}; there is 
a weak minimum in $S$ at $\nucl{12}{C}\nucl{18}{O}/\rm{CO} = 0.022$, 
with an estimated formal error of $\pm0.005$. 

\begin{figure}
   \centering
   \includegraphics[width=8cm]{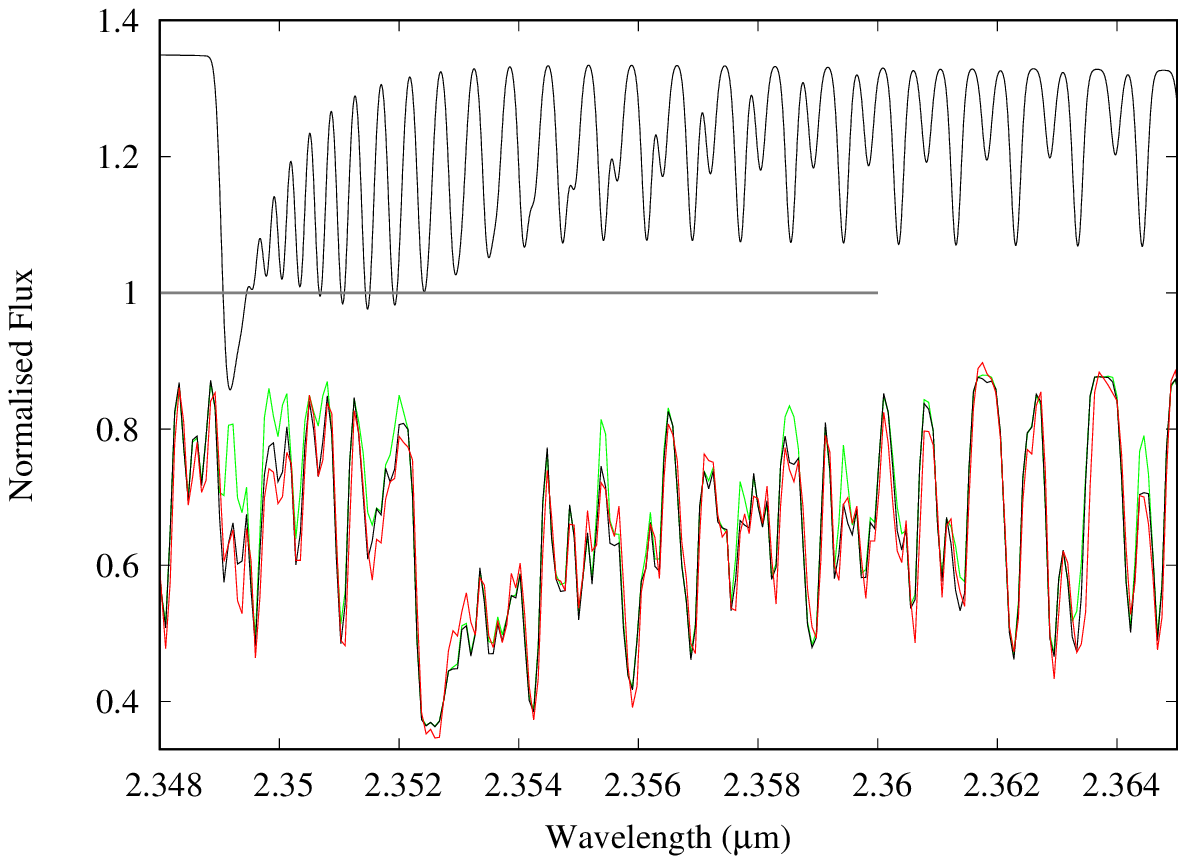}
   \includegraphics[width=8cm]{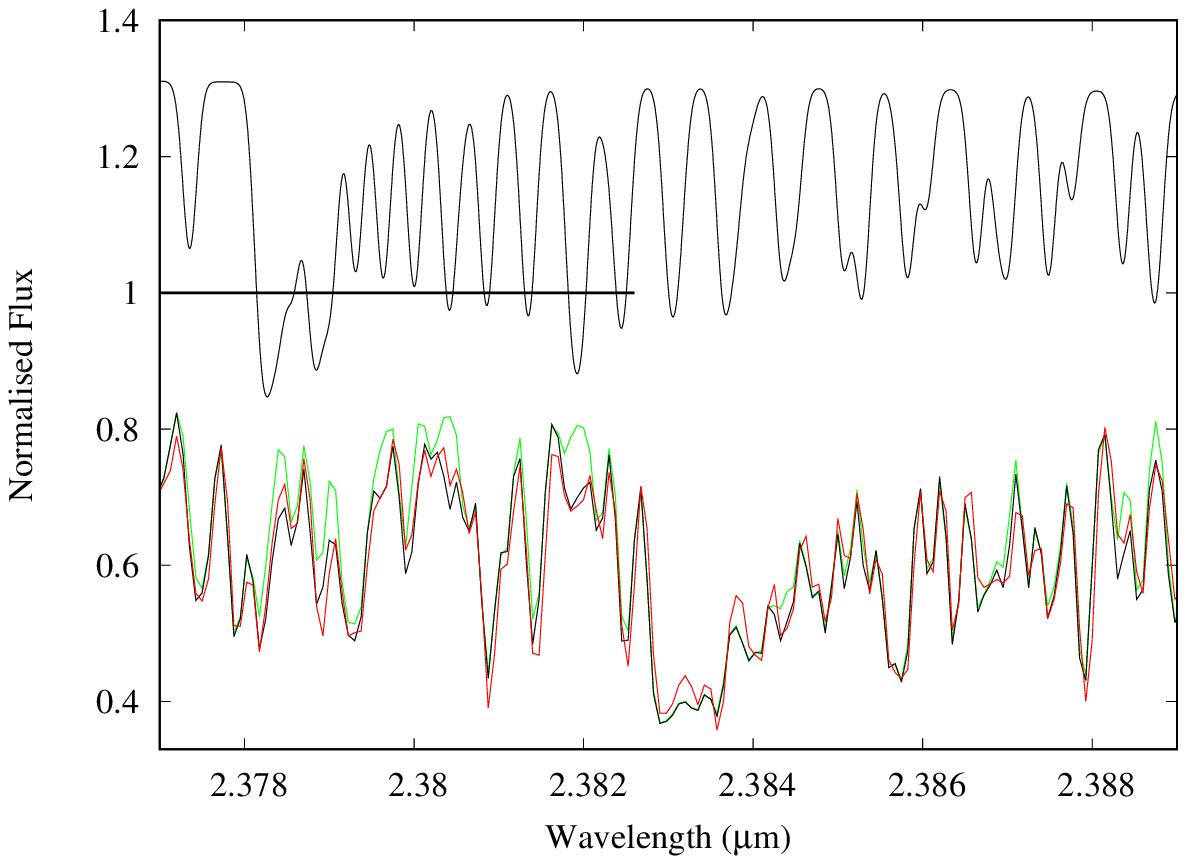}
\caption{Fits to the two selected spectral regions, defined in 
the top panel of Fig.~\ref{18ccs}, 
containing the 0--2 and 1--3 \nucl{12}{C}\nucl{18}{O} bandheads. 
In both panels, black curve is \tcrb, green curve is model
with no \nucl{12}{C}\nucl{18}{O}, red curve is
model with \nucl{12}{C}\nucl{18}{O}/\rm{CO} =0.023.
The dark grey curve at the top of each panel is the 
synthetic \nucl{12}{C}\nucl{18}{O} spectrum.
The horizontal lines at $\mbox{Flux}=1.0$ shows the spectral ranges
used in the minimisation procedure.\label{18cc12}}
  \end{figure}

We then used this result to fine-tune the 
\nucl{12}{C}\nucl{18}{O} abundance by confining our analysis
to the two specific spectral ranges that contain 
a significant contribution from \nucl{12}{C}\nucl{18}{O}; these are 
highlighted in Fig.~\ref{18ccs}, and shown in detail in 
Fig.~\ref{18cc12},
which shows the fit in the region of
the 0--2 and 1--3 transitions of \nucl{12}{C}\nucl{18}{O}\footnote{Note that, in the case of the model spectrum 
with no \nucl{12}{C}\nucl{18}{O}, the model drops below the 
observed spectrum, which does contain \nucl{12}{C}\nucl{18}{O}. 
This apparent anomaly occurs because, in the minimisation procedure,
the absorption by other CO isotopomer(s) has to be increased to attain 
a better fit.}. By fitting the synthetic 
spectra to this restricted wavelength range,
the fit is improved; that this is evident in
seen in the top panel of Fig.~\ref{18ccs}, in which the 
synthetic spectrum without \nucl{12}{C}\nucl{18}{O} (green) provides
a significantly poorer fit to the data.
We find a better-defined minimum $S$ at 
\[ \nucl{12}{C}\nucl{18}{O}/\rm{CO} =0.022\pm0.002 \:\:. \] 
(see Fig.~\ref{18ccs}, bottom panel, and Fig.~\ref{18cc12}). 
We finally find 
\[ \nucl{16}{O}/\nucl{18}{O} = 
N[\nucl{12}{C}\nucl{16}{O}]/N[\nucl{12}{C}\nucl{18}{O}] = 41\pm3.\]

\subsubsection{\nucl{17}{\mbox{\rm O}}/\nucl{16}{\mbox{\rm O}}}

The bandheads of \nucl{12}{C}\nucl{17}{O} nearly
overlap with those of \nucl{12}{C}\nucl{16}{O}
(see Fig.~\ref{ico}), which makes problematic the determination 
of the \nucl{17}{O}/\nucl{16}{O} ratio.
We computed two spectra having relative 
abundances $N(\nucl{12}{C}\nucl{17}{O})/N(\rm{CO}) = 0.0$
and $=0.002$, and computed the ratio of the theoretical fluxes.
We computed $S$ for three spectral ranges 
and we conclude that $N(\nucl{12}{C}\nucl{17}{O})/N(\rm{CO}) < 0.002$.
This result is consistent with the known low abundance of \nucl{17}{O} by 
comparison with other oxygen isotopes in other environments 
\citep[e.g. $\sim4\times10^{-4}$ for the solar photosphere;][]{asplund}.

\subsection{Silicon isotopic ratios}

\begin{figure*}
   \centering
      \includegraphics[width=16cm,height=8cm]{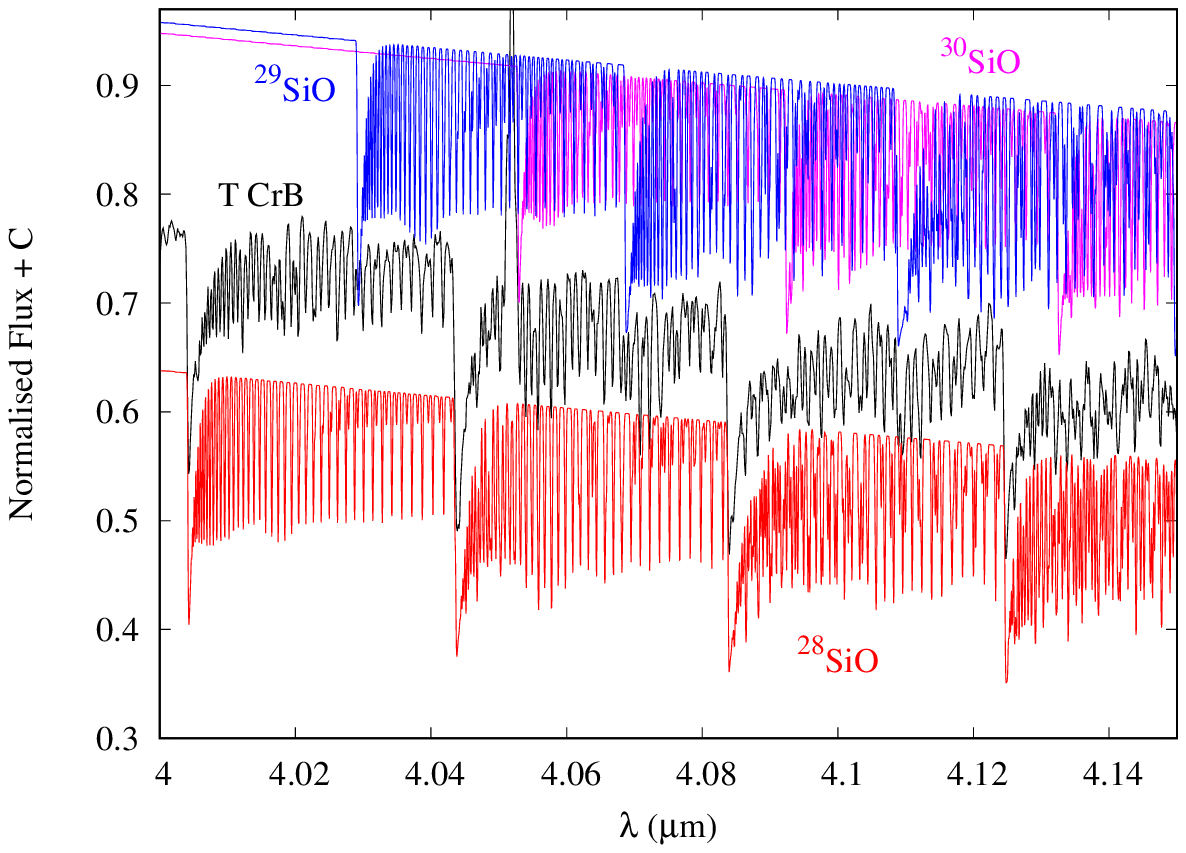}
\caption{Identification of the SiO isotopologues in the region of the first overtone. 
\tcrb\ (black curve):  \nucl{28}{Si}O (red), 
\nucl{29}{Si}O (blue), \nucl{30}{Si}O (magenta).
\label{iSiO}}
  \end{figure*}

The presence of the SiO fundamental band near 8\mic\ in the
spectrum of \tcrb\ was reported by
\cite{evans4}. While some of this was photospheric, there was also a 
contribution from the RG wind, with column density
$2.9\times10^{17}$~cm$^{-2}$. Here we consider the SiO first 
overtone absorption in the wavelength range 4.00--4.15\mic. 
The band heads for 
the various SiO isotopologues, for the various
first overtone transitions in the observed wavelength interval,
are shown in Fig.~\ref{iSiO}, and
listed in Table~\ref{SiOiso}; absorption by SiO dominates 
in the above spectral range. 
As the temperature of the RG wind is $\sim1000$~K \citep{evans4}, 
we may assume that there is no contribution to the SiO first 
overtone band heads from the RG wind.

\begin{table}
\caption{Wavelengths (in \mic, in vacuo) of the SiO first overtone bandheads;
transitions are indicated by $\delta\upsilon$.
From http://www.ukirt.hawaii.edu/astronomy/calib/spec\_cal/sio.html
\label{SiOiso}.}
 \begin{tabular}{cccc}
$\delta\upsilon$   & \nucl{28}{Si}\nucl{16}{O} & \nucl{29}{Si}\nucl{16}{O} 
                  & \nucl{30}{Si}\nucl{16}{O}\\ \hline
2-0 &  4.004 & 4.029 & 4.053 \\
3-1 &  4.043 & 4.069 & 4.092 \\
4-2 &  4.084 & 4.109 & 4.133 \\
5-3 &  4.125 & 4.150 & 4.173 \\
6-4 &  4.166 &  ---  &  ---  \\ \hline
 \end{tabular}
\end{table}

We determined the Si isotopic ratios from the SiO absorption as follows:
\begin{enumerate}
 \item As for carbon, we first determined a self-consistent 
Si abundance and isotopic ratios by varying the abundance of Si in the range
[--0.2, --0.1, 0 +0.1 +0.2], and the relative abundance of 
\nucl{28}{Si}\nucl{16}{O} in the range 0.90--0.97, with step size 0.01.
The process was run iteratively, with model atmosphere and synthetic 
spectra being recomputed for every new Si abundance until we obtained
convergence at $[\mbox{Si}] =-0.1$.
\item We then verified the solution by computing the relative 
isotopic abundances $N(\nucl{28}{Si}\nucl{16}{O})/N(\rm{SiO})$ and 
$N(\nucl{29}{Si}\nucl{16}{O})/N(\rm{SiO})$, varied in the ranges 
[0.80--0.97] and [0.0--0.2], respectively.
 \end{enumerate}

\begin{figure*}
   \centering
  \includegraphics[width=16cm,keepaspectratio]{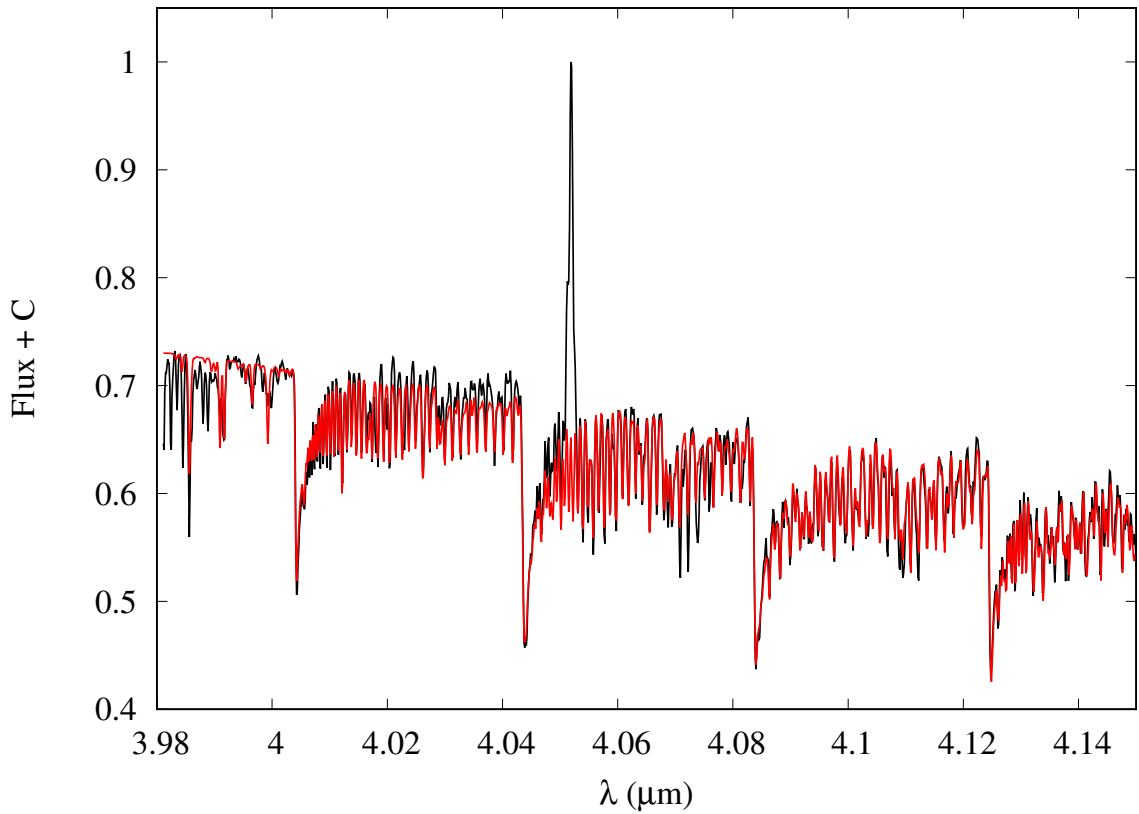}
\caption{Fit (red) to the observed spectrum of \tcrb\ (black) 
across the first overtone SiO bands;  the emission feature is Br-$\alpha$.
\label{2ssio}}
\end{figure*}

The fit to the observed spectrum is shown in Fig.~\ref{2ssio}. 
The first overtone bandheads of \nucl{29}{Si}O at 
4.029\mic, 4.069\mic\ and 4.109\mic\ are clearly present, those of 
\nucl{30}{Si}O less obviously so.
However we must have the constraint that
\[ \nucl{28}{Si}/\mbox{Si} + \nucl{29}{Si}/\mbox{Si} + \nucl{30}{Si}/\mbox{Si} =1 ,\]
as these are the only isotopes of Si.
Minimum $S = 0.332\pm0.011$ was found.
The Si abundances are listed in
Table~\ref{CCOO}, and the isotopic ratios listed in Table~\ref{CCOO2}.
We find 28Si:29Si:30Si=0.86:0.10:0.04, i.e
\nucl{28}{Si}/\nucl{29}{Si} $=8.6\pm3.0$;
 \nucl{28}{Si}/\nucl{30}{Si} $=21.5\pm3.0$.

\begin{table*}
\caption{Logarithmic abundances (by number) of C, O and Si in the 
atmosphere of the RG in \tcrb\ and in the solar photosphere, relative to
$\log{N}(\mbox{H})=12$. \label{CCOO}}
\begin{tabular}{cccccccccc}
       &           &     & \multicolumn{6}{c}{Nova TNR on CO and ONe WDs}   &   \\ 
              &                &   Solar   & \multicolumn{2}{c}{CO$^b$} & \multicolumn{2}{c}{CO$^c$} & \multicolumn{2}{c}{ONe$^d$} & Field RGs\\ \cline{4-9}
 Species    & \tcrb & photosphere$^a$ & 1.25\Msun & 1.35\Msun  & 1.25\Msun & 1.35\Msun & 1.25\Msun & 1.35\Msun & ($\beta$~And, $\delta$~Oph)$^{e}$\\ \hline
C  &  $-3.92$      &  $-3.62$  & $-2.10$ & $-1.78$ & $-1.92$ & $-1.76$ & $-2.74$ & $-2.60$ &  $-3.81$    \\
O  &  $-3.36$      &  $-3.36$  & $-2.65$ & $-3.27$ & $-1.52$ & $-1.86$ & $-1.78$ & $-1.83$ &  $-3.22$    \\
Si &  $-4.69$      &  $-4.54$  & $-3.96$ & $-4.23$ & $-3.41$ & $-3.63$ & $-1.96$ & $-1.52$ &  $-4.65$   \\ \hline
\multicolumn{3}{l}{$^a$\cite{asplund}.} & & & & & &  & \\
\multicolumn{6}{l}{$^b$\cite{starrfield20}. 25:75 mixing during TNR.} & & & & \\
\multicolumn{6}{l}{$^c$\cite{starrfield20}. 50:50 mixing during TNR.} & & & & \\
\multicolumn{6}{l}{$^d$\cite{starrfield09}; I2005A models.}         & & & & \\
\multicolumn{10}{l}{$^{e}$\cite{smith13}. Value for C includes \nucl{12}{C} and \nucl{13}{C};
value for O is for the \nucl{16}{O} isotope only.}  \\
\end{tabular}
\end{table*}

\begin{table*}
\caption{Isotopic ratios (by number) for C, O and Si in the atmosphere of the RG in \tcrb. 
\label{CCOO2}}
\begin{tabular}{cccccccccc}
             &          &      & \multicolumn{6}{c}{Nova TNR on CO and ONe WDs}   &  RG, with initial  \\ 
 Isotope             &       & Solar    & \multicolumn{2}{c}{CO$^b$} & \multicolumn{2}{c}{CO$^c$} & \multicolumn{2}{c}{ONe$^d$} & mass 1\Msun, after \\ \cline{4-9}
 ratio     & \tcrb  & photosphere$^a$ &1.25\Msun&1.35\Msun  &1.25\Msun& 1.35\Msun & 1.25\Msun & 1.35\Msun & 1st dredge-up$^{e}$\\ \hline
\nucl{12}{C}/\nucl{13}{C}    &    $10\pm2$          & 89      & 1.3   & 1.4  & 1.0   & 3.9  &1.8  & 2.8 &26.3 \\
\nucl{16}{O}/\nucl{17}{O}    &   $>500$            &  2625   & 0.1   & 0.17 & 0.61  & 0.63 & 1.3 &0.04 & 2760 \\
\nucl{16}{O}/\nucl{18}{O}    &   $41\pm3$         &   499   & 46    & 9.1  & 1837  & 3562 & 52.9 &1.8 & 556 \\
\nucl{28}{Si}/\nucl{29}{Si}  &   $8.6\pm3.0$       & 19.74   & 26.9  & 29.2 & 23.6  & 31.1 & 5.7 &0.2 & --- \\
\nucl{28}{Si}/\nucl{30}{Si}  &  $21.5\pm3.0$          &  29.75  & 1.9   & 2.2  & 2.8   & 2.3  & 1.8 &0.8 & --- \\ \hline
\multicolumn{3}{l}{$^a$\cite{asplund}.} &&&&&&\\
\multicolumn{6}{l}{$^b$\cite{starrfield20}. 25:75 mixing during TNR.} &&&\\
\multicolumn{6}{l}{$^c$\cite{starrfield20}. 50:50 mixing during TNR.} &&&\\
\multicolumn{6}{l}{$^d$\cite{starrfield09}; I2005A models.}&&&\\
\multicolumn{3}{l}{$^{e}$\cite{karakas}.} &&&&&&&\\
\end{tabular}
\end{table*}

\begin{figure}
   \centering
\includegraphics[width=8cm,keepaspectratio]{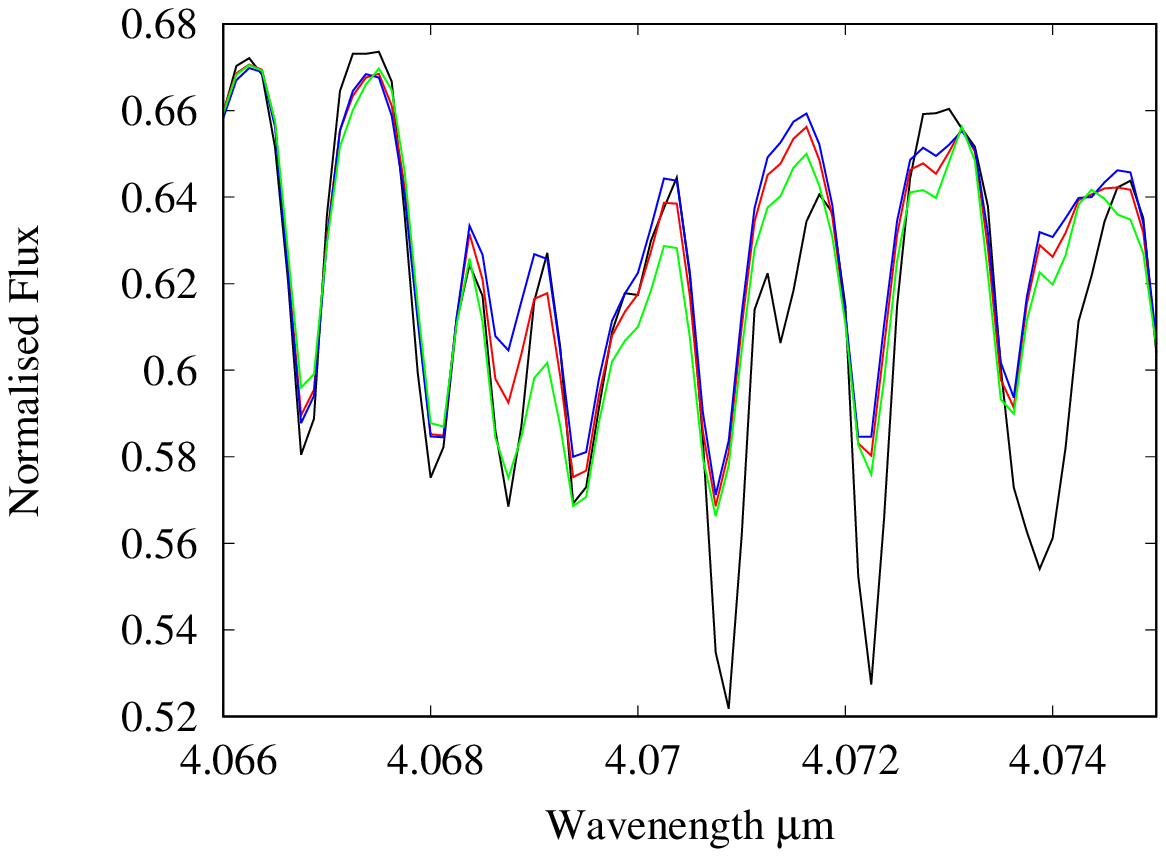}
\caption{Fit to the SiO bands in the 4.066--4.075\mic\ region.
Black curve: \tcrb. 
Red curve: \nucl{28}{Si}/\nucl{29}{Si} = 8.6,
blue curve: \nucl{28}{Si}/\nucl{29}{Si} = 6.8,
green curve: \nucl{28}{Si}/\nucl{29}{Si} = 13.
\label{SiO_err}}
  \end{figure}

\section{Expectation for abundances and isotopic ratios}

\subsection{Elemental abundances}

The deduced abundances for C, O and Si for the RG in \tcrb\
are summarised in Table~\ref{CCOO}, in which the uncertainty in the \tcrb\ 
abundances are $\pm0.1$~dex; we see that 
these elements are somewhat deficient compared to the corresponding solar 
values. The \tcrb\ abundances are similar to those in two field RGs 
\citep[see Table~\ref{CCOO}, in which the uncertaities in the field RG
abundances are typically $\pm0.03$ to $\pm0.08$;][]{smith13}.
However there may be some evidence for a deficiency in C in the RG
in \tcrb. This may be evidence for CNO burning, in which the abundance of
N increases while that of C decreases.

\subsection{Processes in the RG interior}
\label{rgint}
For low-mass ($\sim1$\Msun) stars on the RG branch, first dredge-up 
results in significant changes in the photospheric abundances
\citep[see][for a comprehensive review]{karakas}.
These include a reduction in the C/O ratio (from $\sim0.5$ to 
$\sim0.3$), together with changes in isotopic ratios;
in particlar, \nucl{12}{C}/\nucl{13}{C} declines from the solar 
value ($\sim90$) to $\sim20$, \nucl{16}{O}/\nucl{17}{O} from
$\sim2700$ to $\sim300$, and an increase of \nucl{16}{O}/\nucl{18}{O}
from 524 to 740. 

The predicted decline of the 
\nucl{12}{C}/\nucl{13}{C} ratio to $\sim20$ is exceeded in a variety 
of cases, including RGs in metal-rich clusters (with values as low as $<10$), 
and in RGs in globular clusters. Indeed the two field RGs in
Table~\ref{CCOO} have low \nucl{12}{C}/\nucl{13}{C} ratios
\citep[15 for $\beta$~And, 12 for $\delta$~Oph;][]{smith13}. 
Table~11 of \cite{hinkle16} shows that low values of the 
\nucl{12}{C}/\nucl{13}{C} ratio in evolved stars are unusual, 
but not exceptional \citep[see also][and references therein]{pavl03}.
Such low values of  \nucl{12}{C}/\nucl{13}{C} might point to additional 
mixing after the first dredge-up, such that more material 
processed by CN cycling is mixed into the upper photospheric layers. 
This would result in a decrease in \nucl{12}{C}, \nucl{16}{O} and 
\nucl{18}{O}, while \nucl{13}{C} and \nucl{17}{O} would increase. 

\subsection{Contamination by RN eruptions}
\subsubsection{A cautionary preamble}
During and immediately after a RN explosion, the products of the TNR
reach the RG. The fate of these products however is unclear,
and the issue of contamination of the RG by a TNR is a complex one, 
even if the products remain on the surface rather than sweep past the 
RG without contaminating it. If the products do pollute the RG surface,
only one hemisphere of the RG feels the impact of the ejecta, and 
whether or not the contaminants remain on the RG surface or are convected into
the RG interior is an open question. And even if the pollutants remain
on the surface, the time it takes for them to propagate over the entire 
RG surface, and whether contamination from previous RN eruptions remains
on the surface, is also unknown. Furthermore, any TNR products
deposited on the RG surface would likely eventually be removed by the RG wind.

Given all these caveats and uncertainties, it is unclear 
to what extent we can directly 
compare the surface abundances of the RG in \tcrb\ with the products 
of a TNR. However we summarise here the likely products of the TNR
(Tables~\ref{CCOO} and \ref{CCOO2}),
and in our discussion in Section~\ref{implications} we assume for our
present purposes that the TNR products pollute the entire RG surface.

\subsubsection{The TNR}
\citeauthor{starrfield09} have undertaken a study of the 
TNR on the surfaces of ONe \citep{starrfield09} and
CO \citep{starrfield20} WDs. As the WD in the \tcrb\ system
seems to have mass close to the Chandrasekhar limit
\citep{shahbaz2,kennea,shara}, we confine our discussion
of these papers to the products of TNR on the high mass end of WDs.

For TNRs on CO WDs, \cite{starrfield20} include mixing of 
WD and accreted material, the latter assumed to have solar composition;
they consider mixes of 25\%WD:75\%accreted, and 50\%WD:50\%accreted. 
The expected C, O, Si abundances, and corresponding isotopic ratios for 
these cases are included in Tables~\ref{CCOO} and \ref{CCOO2} for WD masses
1.25\Msun\ and 1.35\Msun. The Tables include the same information
for TNRs on the corresponding ONe WDs \citep{starrfield09}.

\cite{starrfield09,starrfield20} find that TNRs on massive WDs of both
CO and ONe types produce large overabundances of \nucl{13}{C}
and \nucl{17}{O}, while overabundances of \nucl{18}{O}
are produced on ONe WDs, and on CO WDs if 75\% of the accreted material
partakes in the TNR. However only TNRs on ONe WDs 
synthesise substantial amounts of Si, with both 1.25\Msun\ and 
1.35\Msun\ ONe WD producing overabundances of \nucl{28}{Si}, 
\nucl{29}{Si} and \nucl{30}{Si}. 

\section{Implications for \tcrb}
\label{implications}

We first note that Si in the photosphere of the RG in 
\tcrb\ is not overabundant, and is close to that in the solar
photoshere and in field RGs (see Table~\ref{CCOO}). Thus if the RG is 
polluted by the products of a TNR then the latter is unlikely to have 
occurred on an ONe WD.

However our deduced $\nucl{12}{C}/\nucl{13}{C} = 10$ is low, and could 
be the outcome of either the additional mixing discussed
in Section~\ref{rgint}, or of contamination 
of the RG photosphere by material from previous RN outbursts. 
As summarised in Table~\ref{CCOO2}, a TNR can 
result in a \nucl{12}{C}/\nucl{13}{C} ratio close to unity
\citep{starrfield09,starrfield20}: the ratio in the RG in \tcrb\
does not discriminate between these two scenarios.
Similar values of \nucl{12}{C}/\nucl{13}{C} to that in \tcrb\
have also been found in Miras \citep*[e.g.][]{hinkle16}.

In a 1\Msun\ RG, a ratio of \nucl{16}{O}/\nucl{17}{O}
$\sim2700$ is predicted after the first dredge-up 
\citep{karakas}, and as we are able to place only a lower limit 
on the  \nucl{16}{O}/\nucl{17}{O} ratio ($> 500$) in \tcrb,
we are unable to use this to determine whether the 
low \nucl{12}{C}/\nucl{13}{C} ratio is the result of
contamination from a RN eruption, or additional mixing following
first dredge-up. The \nucl{17}{O} abundance in \tcrb\ may 
be normal for a RG, and is not inconsistent with the values
found in Miras \citep{hinkle16} and in some field RGs (see 
Table~\ref{CCOO}).

\begin{figure}
   \centering
         \includegraphics[width=8cm,keepaspectratio]{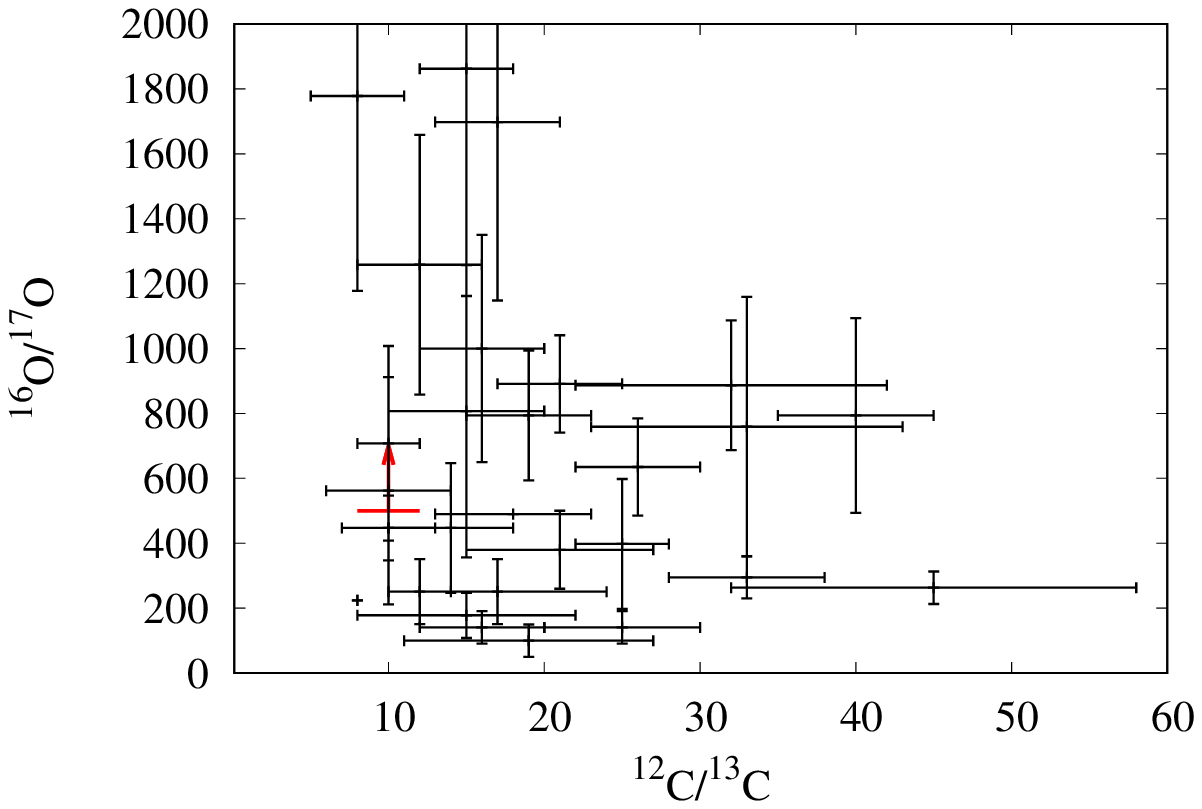}
               \includegraphics[width=8cm,keepaspectratio]{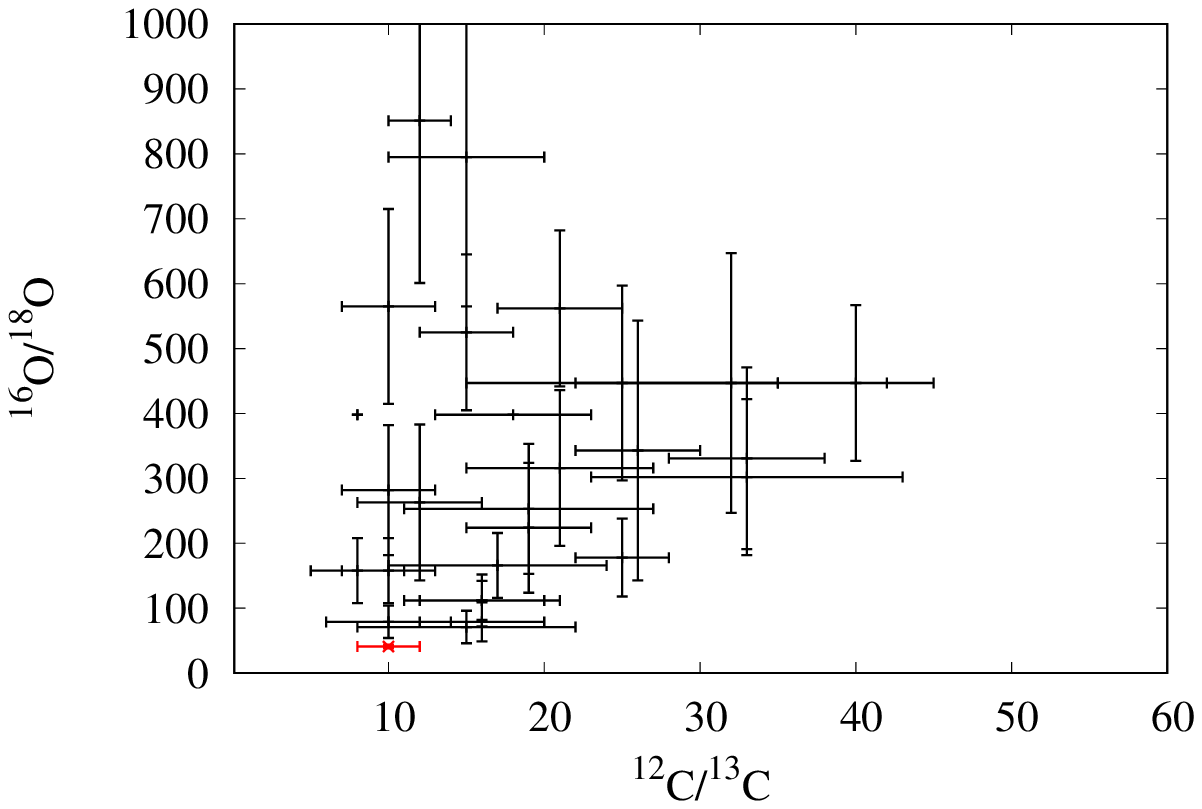}
      \includegraphics[width=8cm,keepaspectratio]{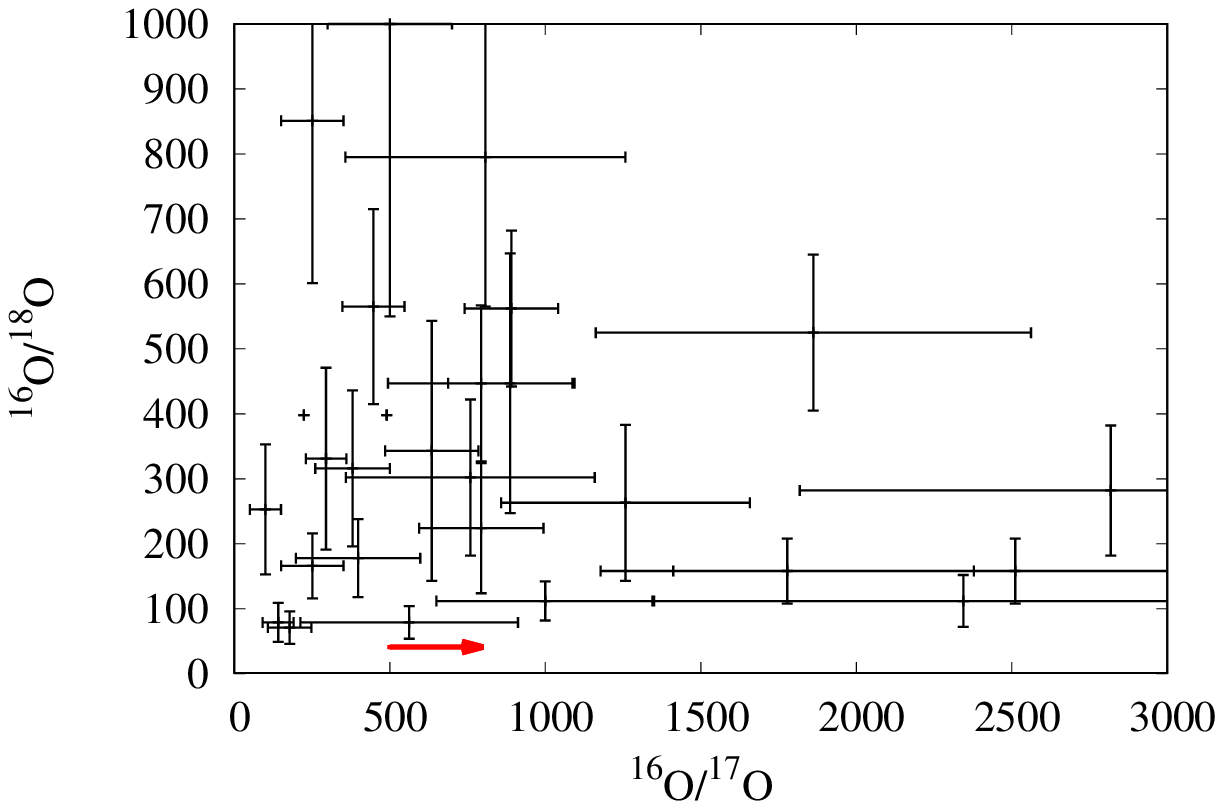}
\caption{Comparison of the C and O isotopic 
rations in \tcrb\ with those in field RGs, as measured by Hinkle et al. (2016).
Data for \tcrb\ are in red. \label{Hcomp}}
  \end{figure}

On the other hand, \nucl{18}{O} appears to be overabundant in \tcrb, 
as suggested by the deduced ratio of \nucl{16}{O}/\nucl{18}{O} 
$\sim41$. The predicted value is $\sim550$ for a 
1\Msun\ RG, both before and after first dredge-up, while in the 
solar neighbourhood, its value is $\sim500$ 
\citep*[][and references therein]{harris84,geiss02,scott06}.
The anomalous nature of the C and O isotopic 
ratios for \tcrb\ are highlighted in Fig.~\ref{Hcomp},
in which we have plotted the isotopic ratios for \tcrb\ and
the M stars in \cite{hinkle16}'s compilation (omitting carbon
and S~stars). The isotopic ratios for \tcrb\ are clear outliers in the 
isotopic ratio planes.

Large excesses of \nucl{18}{O} have been found in highly evolved objects, 
for example in several hydogen-deficient and R~CrB stars, with the 
\nucl{16}{O}/\nucl{18}{O} ratios being close to, and in some 
cases less than, unity \citep{clayton07}. However, the events that have 
been proposed to produce these unusual values are WD mergers, rather than novae.

We have checked the dependence of the
C and Si abundances on effective 
temperature, and find that a change from 3600~K to 3500~K changes 
[C] by 0.1, and [Si] by 0.01.
Lower values of microturbulent velocity than the 3\vunit\ we have assumed
lead to significantly larger values of our fitting parameter $S$
(e.g. $S=5.77$ for 1\vunit, and $S=3.14$ for 2\vunit, all other
parameters being held fixed). We further find that using $\logg=0$ 
has negligible effect on $S$.

We also note that \cite{pavl20a} recently obtained for the isolated giant
Arcturus (K2 III) isotopic ratios \nucl{12}{C}/\nucl{13}{C} $=10\pm2$ 
(similar to that found in \tcrb) and \nucl{16}{O}/\nucl{18}{O} 
$= 2000 \pm 500$ \citep{pavl20a}. The Arctutus results 
were obtained using exactly the the same procedures as those used here,
demonstrating that our results are not a feature of the methodology.

In all these cases the 
evolutionary histories are of course completely different from that of \tcrb.
Furthermore, the combined evidence of the 
\nucl{16}{O}/\nucl{18}{O} and \nucl{16}{O}/\nucl{17}{O} 
ratios in \tcrb\ is extremely puzzling: together they are not 
compatible with either production in a TNR or in dredge-up.

\section{Conclusions}

We have presented near-IR spectroscopy of the RN \tcrb\ 
and we find that, for the RG in this system:
\begin{enumerate}
 \item the \nucl{16}{O}/\nucl{18}{O} ratio is low, $\sim41\pm3$;
 \item the \nucl{12}{C}/\nucl{13}{C} ratio is $10\pm2$;
 \item the \nucl{16}{O}/\nucl{17}{O} ratio is $>500$;
 \item the \nucl{28}{Si}/\nucl{29}{Si} ratio is $8.6\pm3.0$;
\item the \nucl{28}{Si}/\nucl{30}{Si} ratio is $21.5\pm3.0$.
  \end{enumerate}
Taken together, the C and O isotopic ratios are puzzling, and 
confirmation by obtaining and modelling higher resolution spectroscopy (e.g., $R\sim60,000$) 
is highly desirable.
Observations of other evolved stars
(e.g. field RGs, Miras) at such high resolution have resulted 
in modelling individually resolved CO lines 
\citep*{harris84,abia12,hinkle16}\footnote{We note that, while 
\cite{harris84} used the CO fundamental band transitions at 4.67\mic,
\cite{hinkle16} used the CO first overtone transitions.}. 
The results for the \nucl{16}{O}/\nucl{18}{O} ratio from the
latter two studies match very closely for all the stars, with 
\nucl{16}{O}/\nucl{18}{O} values between 
400 and 500.

We conclude that \tcrb\ shows evidence for a large overabundance of 
\nucl{18}{O}. Whether this is connected with the 
RG in \tcrb\ being a member of a binary system that periodically
undergoes TNR is open to debate. But in addition to the need for
verification with high resolution spectroscopy, it is also 
highly desirable that the extent to which nova ejecta -- for both
classical and recurrent novae -- pollute the atmosphere of the 
secondary is examined in detail. If the contamination is long-term
(i.e. comparable with the inter-outburst period), the implications 
for the composition of the material accreted on to the WD will
be substantial. Understanding this process will be important to 
understanding the evolution of nova systems.

\section*{Acknowledgments}

We thank the referee for their helpful comments
on an earlier version of this paper. AE thanks Professor Sergey Yurchenko,
UCL, for helpful information about the ExoMol data.

This paper is based on observations obtained for Fast Turnaround 
programme GN-2019A-FT-207 of the Gemini Observatory, which is operated by 
the Association of Universities for Research in Astronomy, Inc., under
a cooperative agreement with the NSF on behalf
of the Gemini partnership: the National Science Foundation (United States),
National Research Council (Canada), CONICYT (Chile), Ministerio de Ciencia,
Tecnolog\'{i}a e Innovaci\'{o}n Productiva (Argentina), Minist\'{e}rio da
Ci\^{e}ncia, Tecnologia e Inova\c{c}\~{a}o (Brazil), and Korea Astronomy
and Space Science Institute (Republic of Korea).

YP's work was funded as part of the routine financing program 
for institutes of the National Academy of Sciences of Ukraine.
DPKB is supported by a CSIR Emeritus Scientist
grant-in-aid and is  being hosted by the Physical Research
Laboratory, Ahmedabad.
RDG acknowledges support from NASA and the United States Airforce.
SS acknowledges partial support to ASU from HST and NASA.

\section*{Data availability}
The raw data in this paper are available from the Gemini Observatory Archive,
https://archive.gemini.edu/

\label{lastpage}
\end{document}